\RequirePackage{fix-cm}
\documentclass[smallextended]{svjour3}       
\pdfoutput=1
\smartqed  
\usepackage{graphicx}
\usepackage[square,sort,comma,numbers]{natbib}
\usepackage{aas_macros}

 \journalname{Experimental Astronomy}
\begin{document}

\title{Gravitation And the Universe from large Scale-Structures
}
\subtitle{The GAUSS mission concept\\ 
Mapping the cosmic web up to the reionization era}


\author{Alain Blanchard         \and
 Aubourg Éric 
           \and
Philippe Brax 
           \and
Francisco J. Castender 
           \and
 Sandrine  Codis 
           \and
Stéphanie Escoffier 
           \and
Fabien Dournac 
           \and
Agnès Fert\'e
           \and
Fabio Finelli 
           \and
Pablo Fosalba 
           \and
Emmanuel Gangler 
           \and
A Gontcho Satya Gontcho 
           \and
Adam Hawken 
           \and
Stéphane Ili\'c 
           \and
Jean-Paul Kneib 
           \and
Martin Kunz 
           \and
Guilhem Lavaux 
           \and
Olivier Le Fèvre 
           \and
Julien Lesgourgues 
           \and
Yannick Mellier 
           \and
Jérémy Neveu 
           \and
Yann Rasera 
           \and
Cécile Renault 
           \and
Marina Ricci 
           \and
Ziad Sakr 
           \and
Norma G. Sanchez 
           \and
Isaac Tutusaus 
           \and
Safir Yahia-Cherif 
}


\institute{
Alain Blanchard \at  IRAP, CNRS, Toulouse, France\\\email{alain.blanchard@irap.omp.eu}
           \and
Éric Aubourg \at APC, CNRS, Paris, France\\
           \and
Philippe Brax \at  IPhT, CEA, Saclay, France\\
           \and
Francisco J. Castender \at  ICE-CSIC, IEEC, Barcelona, Spain\\
           \and
 Sandrine  Codis \at  IAP, CNRS, Paris, France\\
           \and
Stéphanie Escoffier \at  CPPM, CNRS, Marseille, France\\
           \and
Fabien Dournac \at  IRAP, CNRS, Toulouse, France\\
           \and
Agnès Fert\'e \at  JPL, Pasadena, USA\\
           \and
Fabio Finelli \at  INAF, Bologne, Italy\\
           \and
Pablo Fosalba \at  ICE-CSIC, IEEC, Barcelona, Spain\\
           \and
Emmanuel Gangler \at  LPC, CNRS, Clermont-Ferrand, France\\
           \and
A Gontcho Satya Gontcho \at  University of Rochester, USA\\
           \and
Adam Hawken \at CPPM, CNRS, Marseille, France\\
           \and
Stéphane Ili\'c \at   LERMA, CNRS, Paris, France\\
           \and
Jean-Paul Kneib \at  EPFL, Lausanne, Switzerland\\
           \and
Martin Kunz \at  University of Geneva, Switzerland\\
           \and
Guilhem Lavaux \at  IAP, CNRS, Paris, France\\
           \and
Olivier Le Fèvre \at  LAM, CNRS, Marseille, France\\
           \and
Julien Lesgourgues \at  TTK, RWTH Aachen University, Germany\\
           \and
Yannick Mellier \at  IAP, CNRS, Paris, France\\
           \and
Jérémy Neveu \at LAL, CNRS, Orsay, France\\
           \and
Yann Rasera \at  LUTH, CNRS, Meudon, France\\
           \and
Cécile Renault \at  LPSC, CNRS, Grenoble, France\\
           \and
Marina Ricci \at  LAPP, CNRS, Annecy, France\\
           \and
Ziad Sakr \at  USJ, Beirut, Lebanon\\
           \and
Norma G. Sanchez \at  LERMA, CNRS, Paris, France\\
           \and
Isaac Tutusaus \at  ICE-CSIC, IEEC, Barcelona, Spain\\
           \and
Safir Yahia-Cherif \at  IRAP, CNRS, Toulouse, France\\
}

\date{Received: date / Accepted: date}

\maketitle

\begin{abstract}
Today, thanks in particular to the results of the ESA Planck mission, the concordance cosmological model appears to be the most robust to describe the evolution and content of the Universe from its early to late times. It summarizes the evolution of matter, made mainly of dark matter, from the primordial fluctuations generated by inflation around $10^{-30}$ second after the Big-Bang to galaxies and clusters of galaxies, 13.8 billion years later, and the evolution of the expansion of space, with a relative slowdown in the matter-dominated era and, since a few billion years, an acceleration powered by dark energy. But we are far from knowing the pillars of this model which are inflation,  dark matter and  dark energy. Comprehending these fundamental questions requires a detailed mapping of our observable Universe over the whole of cosmic time. The relic radiation provides the starting point and galaxies draw the cosmic web. JAXA's LiteBIRD mission will map the beginning of our Universe with a crucial test for inflation (its primordial gravity waves), and the ESA Euclid mission will map the most recent half part, crucial for dark energy. The mission concept,  described in this White Paper, GAUSS, aims at being a mission to fully  map the cosmic web up to the reionization era, linking early and late evolution, to tackle and disentangle  the crucial degeneracies persisting after the Euclid era between dark matter and inflation properties, dark energy, structure growth and gravitation at large scale.\\

\end{abstract}

\section{Science motivations}
\label{intro}
The questions regarding very high energy physics (far beyond the energies accessible to terrestrial accelerators), and the origin of the acceleration of the expansion of the Universe are very important and timely questions of both fundamental physics and cosmology. Cosmological and astronomical observations will provide the essential information, if not all the information needed to answer these questions. The study of the distribution of matter at very large scales, allowed by large surveys of galaxies, complemented by the observation of the properties of the cosmic microwave background (CMB) radiation, are the primary probes of this new physics, which remains to be discovered and understood. \\
 
The most direct and simplest explanation of the origin of dark energy which is compatible with all the available data is the cosmological constant. However, it needs fine-tuning and as a vacuum energy suffers from the so called "cosmological constant problem", i.e. the theoretical vacuum energy value estimated from quantum particle physics is about $10^{122}$ times the observed value \citep{1989RvMP...61....1W}. Such a huge difference between the two values could physically correspond to two different vacuum states of the universe, namely at two different epochs of its evolution: one being the classical very large universe today, the other being the very early quantum cosmological vacuum. The low value of $\Lambda$ or vacuum energy density today corresponds to a large-scale low-energy diluted universe essentially dominated by voids and supervoids as the set of large-scale observations concordantly and independently shows. On the other hand, the high quantum value of $\Lambda$ could correspond to the high energy and highly dense, small-scale very early quantum vacuum.\\

A different proposal for the origin of the present acceleration supposes a modification of the General Relativity theory (GR) of gravity by breaking one of the assumptions of Lovelock's theorem \citep{1971JMP....12..498L}, which states that the only possible equations of motion which contain only second derivatives in 4-dimensional space-time from a scalar Lagrangian density satisfying Lorentz invariance are the Einstein field equations. One common modification is to include a scalar field coupled to the curvature tensor terms in these equations through the Horndeski Lagrangian \citep{1974IJTP...10..363H}. This includes all possible introductions of scalar, vector, and tensor degrees of freedom. Among these theories, most of the additional tensor degrees of freedom have been significantly constrained by the discovery of a gravitational wave source and its optical counterpart \citep{2017PhRvL.119p1101A}. Moreover, recent analyses of Integrated Sachs-Wolfe effect (ISW)-galaxy cross-correlation data \citep{2017JCAP...10..020R} have also disfavored other derivative couplings of the cubic order as well as beyond Horndeski \citep{2014JCAP...10..071L} higher order additional theories. Theories yet to be excluded are those with linear scalar self-interactions and minimal/derivative couplings to gravity, such as dark energy models based on a scalar field with a potential (quintessence) \citep{1988PhRvD..37.3406R} or with a non canonical kinetic term (k-essence) \citep{1999PhLB..458..209A}. Other modifications of gravity are still also viable, modifying the remaining assumptions such as models with extra dimensions (e.g. Kaluza-Klein type \citep{1921SPAW.......966K,2009NuPhB.821..467K}), or those that relax the Lorentz invariance (e.g. Horava-Lifshitz gravity \citep{2009NuPhB.821..467K}). But these can all be distinguished from each other by either a different evolution of the background expansion and/or a different cosmic growth history. Some of these models of modified gravity could be used to incorporate both late time acceleration and inflation at early times. All these models could be tested with CMB cross-correlation observations, however the latter probe, unlike the distribution of large scale structure, only growth and expansion at the time of recombination. Next generation CMB experiments exploring high multipole modes at small scales will need to account for the foreground distribution of large-scale structure. \\

At the moment, no favoured model of dark energy seems to emerge implying that an effective description of this phenomenon using a finite and well-chosen set of parameters should be employed. \\

Among these phenomenological tools, the measure of the expansion $H(z)$, and growth rate of structure will provide probes for dark energy over cosmic times and will indirectly probe very high energy cosmic origins. Moreover, the distribution of matter, traced by galaxies and dark matter through the effects of gravitational lensing and cross correlations between the former and latter distribution, offers an even more powerful tool to better understand the characteristics of cosmic expansion and the growth rate of structure. The reason comes from the fact that the lensing distribution measurements are sensitive to the sum of the two gravitational potentials $\Phi + \Psi$ \citep{1980PhRvD..22.1882B} while galaxy clustering, being non-relativistic, is sensitive to the Newtonian potential $\Psi$ only. Within GR, the two are equal, while in modified gravity models there can be a shift called the slip parameter. Therefore, the combination of the two observables probes the relation between the two gravitational potentials. Moreover, the precise estimation of weak lensing and galaxy clustering statistics will help disentangle degenerate effects from the aforementioned extensions to the current concordance model (or $\Lambda$-CDM, named from its main components, dark energy and cold dark matter), i.e., the slip, growth, and expansion of the universe, through the determination of the $E_g$ factor \cite{2006SPIE.6269E..15M} which encapsulates three related observables: the Hubble parameter, the galaxy-velocity and galaxy-lensing cross spectrum.  \\

This distribution of matter is also a diagnostic test of the physics of the early universe: thus the presence of massive neutrinos could be highlighted for a range of masses inaccessible to experiments in accelerators and potentially better constrained than from CMB experiments of the same calibre \citep{2016PhRvD..94h3522G}. This is due to the fact that the dominant effect of massive neutrinos is the suppression of the growth of structure on relatively small scales, while neutrino masses of a few tenths of eV are still relativistic at the time of photon decoupling and affect the CMB at the background level. There are two effects that massive neutrinos could have on CMB anisotropies \citep{2005NuPhS.145..313H,2006PhR...429..307L}: one through the early ISW, which is roughly ten times smaller than the depletion in the small-scale matter power spectrum; and the other through the late ISW effect which is difficult to measure due to cosmic variance. Moreover, the fact that the suppression of the growth by massive neutrinos is time-dependent leads to increasingly better constraints from the extension of the redshift coverage of large scale structure surveys.\\

The distribution of galaxies and large-scale structures is also a diagnostic tool of the physics of inflation. In these scenarios this distribution is indeed generated via the fluctuations of the metric produced during a high energy phase transition in the early universe, perhaps at energies on the order of $ 10^{15} $ GeV. 
A favoured path of testing inflation would be the detection of B modes of the polarization of the cosmological background. The amplitude of these modes, from the tensor to scalar ratio $ r $, is mandated to be non zero whatever the inflation model, with the most recent CMB data having provided $ r < 0.04-0.07 $ \citep{2020arXiv201001139T}.  Besides just improving such constraints, a detection, if any, would require - and justify - a dedicated mission. \\

The possible presence of a very weak non-gaussian signal is another prediction of inflation, for which a minimal level is expected and would be accessible to surveys that cover a large enough volume of the Universe. Untill now, the different sets of robust and independent cosmological data clearly favor minimal inflationary models for which the amount of non-gaussianity indicated by the parameter $ f_{NL} $ is very small. e.g., the most recent data provides a constraint of $ f_{NL} < 6$  \cite{Planck:2018xyz}. Non-minimal single field models, and especially multi-field models, produce large non-gaussianity $ f_{NL} >~ 1 $ \cite{2017PhRvD..95l3507D} , which may be considered the threshold to detect and constrain. Theoretically, the predictive minimal models, providing the whole set of inflationary observables like $ (n_s, r, f_{NL}) $ and adiabaticity, indicate $f_{NL}$ to be of the order $10^{-2}$, more precisely $f_{NL} \sim (1 / N_e) \sim 0.02$, $N_e$ being the inflation number of e-folds \cite{2006PhRvD..73b3008B, 2017PhRvD..95l3507D}. Overall, the direction in which both the combined robust observational data and predictive theory are pointing is towards a very small amount of primordial non-gaussianity.\\

Within the horizon at 2035, it is important to keep in mind that in the primordial phases of the Universe, besides inflation and its GUT energy scale at time scales of $ 10^{-32} $ sec, there is room for higher energy scales at earlier times which are of the order of the Planck fundamental scale $10^{19}$ GeV at $10^{-44}$ sec and beyond, i.e., the so-called trans-Planckian regime. This quantum phase and its late imprints is a targeted field of study in quantum unification theories, gravitation and cosmology. The understanding of dark energy, whether within or outside of a standard concordance model description, (namely the cosmological vacuum energy, cosmological constant, or dynamical dark energies for instance), is at the center of these studies. Deviations from the classical $\Lambda$-CDM model description if measured in the expansion rate and/or the growth of structures would thus impact both the cosmological community and the wider context of fundamental physics, including string theory.

\section{Probes}
\label{probes}
Different probes are used to study primordial inflation, cosmic acceleration and the growth of  structures. We propose to make precise measurements of the cosmic web, up to unprecedented redshift, with a high density of sources, to encompass at once matter and space evolutions. \\

The design of the GAUSS mission concept is optimized for the combination of probes called “$3 \times 2$pt”. The combination of weak lensing (WL) and galaxy clustering (GC) is particularly powerful at simultaneously constraining the expansion of the universe, by measuring distances thanks to standardized objects or  typical features (like the baryonic acoustic oscillation (BAO) scale), and probing the matter power spectrum, especially its time evolution thanks to measurements in redshift bins, referred to as tomography. Such a survey, if deep and dense enough, should allow us to safely separate the dark energy equation of state, modified gravity effects (if any), and the total mass of neutrinos. Proper access to large and small scales, in time and in space, will permit us to break the degeneracies from which the experiments of the Cosmic Vision generation will suffer.\\

Current galaxy surveys, e.g. DES \cite{2018PhRvD..98d3526A} or KIDS \cite{2018MNRAS.474.4894J}, already combine WL with GC measurements to constrain cosmological parameters and cosmological models. This combination increases the constraining power and is especially powerful at breaking the degeneracy between cosmology and galaxy bias as well as the impact of intrinsic alignments. We describe these probes in the following and also introduce related probes and statistics that are only rarely used in current cosmological analysis but which we will be able to measure precisely with a deep spectro-imager like GAUSS, thus allowing a greater exploration of the universe’s evolution and contents. \\

The clustering is the measure of the 2-point correlation function - even if higher order may also be computed - between the position of tracers of the underlying dark matter structures as compared to the one of a random field. It therefore tells us how clustered is the universe, thus giving us information on the universe evolution and contents. The GAUSS mission concept aims at using galaxies and galaxy clusters as tracers.\\

Gravitational lensing is the deformation of light paths by the presence of a mass acting as a lens by curving space-time. In the regime of diffuse massive structures being the lens, this effect is referred to as 'weak' lensing as it can be detected only through statistical measurements. Indeed the induced ellipticity is at most one order of magnitude lower than the intrinsic ellipticity of the galaxy shapes. Weak lensing affects light of sources or diffuse background. Here we focus on the WL of background galaxies, the 'sources', which requires mapping the sky with a galaxy survey. \\

WL changes the shape of background galaxies, referred to as cosmic shear, and the size of source galaxies, an effect called magnification. For the past two decades, cosmic shear has been the main measurement of WL, e.g. \cite{Kaiser:2000if} or \cite{2000A&A...358...30V}. It is estimated by measuring shapes, and in particular ellipticities of source galaxies. The main derived statistics is then the 2-point correlation function of the ellipticities in redshift bins. This shear statistic is directly linked to the matter power spectrum. Measuring shear in tomography is therefore a great probe of the growth of structure and used in current and forthcoming cosmological surveys to constrain $\Lambda$-CDM model parameters and test beyond $\Lambda$-CDM models. The magnification modifies size and magnitude of galaxies because it changes size while conserving surface brightness. It also changes the number density of galaxies as some background galaxies become detectable thanks to the flux enhancement. WL measurements using magnification have been done, e.g. with DES \cite{2018MNRAS.476.1071G}, through these different effects. Magnification and shear probe the same underlying dark matter distribution but are prone to different systematics, so they will help constrain cosmology better. \\

$3\times 2$pt refers to the combining of three 2-point correlation functions: the clustering, the cosmic shear and the correlation of galaxies and the shear (referred to as galaxy-galaxy lensing). In the end, combining $3\times 2$pt and magnification in tomography will enable a precise measurement of the growth of structure and improving our understanding of the origin of the current cosmic acceleration.  \\

Also, other statistics than the 2-point correlation functions can add valuable information. For instance, cosmology can be constrained using peaks in the convergence maps \cite{2019PhRvD..99d3534V} which gives access to non-Gaussian information or by measuring 3-dimensional WL; its relation with constraints on modified gravity have been shown in \cite{2016PhRvD..93j3524P}. \\

Ultimately, precise WL measurements rely on exquisitely imaging galaxies and on the correct estimation of redshift, both of which are enabled by a spectro-imager. The expected increase of galaxy densities provided by GAUSS, with respect to Euclid, will allow us to measure the matter power spectrum in detail along the whole of cosmic time, from large scales (thanks to a very large sky coverage), down to the galaxy cluster scale (thanks to a high galaxy density). This, in turn, should allow us to disentangle most of the degeneracies between the growth of structure, gravitation at large scales, and the total mass of neutrinos.\\

Naturally, some astrophysical systematic effects will complicate the work. For instance, galaxy peculiar velocities give rise to distortion in the redshift space distribution of galaxies relative to their real space distribution. Furthermore, the intrinsic alignment of galaxies, observed even if not yet understood, affects the measurement of cosmic shear. Once again, the tremendous statistics should allow the study of these interesting effects as well  and to smooth them for cosmology.\\
   
While the $3\times 2$pt and the 3D matter power spectrum are the most promising probes for cosmology with such a survey, galaxy clusters, quasars, and strong lensing could also be powerful probes. Dedicated studies of these probes can be made from the same raw data. In addition to providing meaningful astrophysical results, by providing an internal check on the coherence of all cosmological probes in the same survey, these studies will permit us to control systematic effects. The use of type Ia supernovae would require a specific observational strategy as light curves are mandatory. This probe may suffer from many astrophysical systematic effects at very high redshift and will be largely exploited at low and intermediate redshift by the Rubin Observatory (previously known as the LSST) in conjunction with spectra measured by the ATLAS project \cite{2019BAAS...51c.508W}. Therefore the GAUSS mission concept has not been envisioned to include this probe.\\

Note that the other quantities, $E_g$ especially designed to separate modified gravity from expansion and growth behaviors (cf. section I), and $f_{NL}$ aiming to follow the primordial fluctuations properties will be highly relevant to a GAUSS-like survey. These quantities are emerging for galaxy surveys \cite{Ghosh:2018ijm}, \cite{Castorina:2018zfk}, \cite{Castorina:2019wmr} and no precise predictions regarding their constraints are presented in this White Paper. \\

$E_g$ is an observable aimed at being a model independent gravitational consistency check \cite{2007PhRvL..99n1302Z}. It can be performed by using the same set of galaxies that serve to trace non-relativistic gravitationally-driven motion, and as foreground lenses for probing the relativistic deflection of light from background sources. In this way, it could be ascertained whether the relative amplitude of these two effects, driven by the same underlying density perturbations traced by the lenses, assuming an expansion for a given set of cosmological parameters, is consistent with the prediction of GR. 
It is carried out by cross-correlating lens galaxies to both the surrounding velocity field using redshift spectral distortion and to the shear of background galaxies using galaxy-galaxy lensing \cite{2016MNRAS.456.2806B}. While galaxy-galaxy lensing, galaxy clustering, and galaxy redshift distortions are strongly sensitive to the galaxy bias and to the amplitude of the matter perturbations, however, the combination of these quantities in $E_g$ is such that both nuisance parameters cancel out. Also, CMB lensing has been proposed 
as a more robust tracer of the lensing field for constructing $E_g$ at higher redshifts and at large scales while avoiding intrinsic alignments. 
Few constraints have been put on $E_g$ using these two methods and on several datasets for sparse redshift ranges \cite{Pullen:2015vtb} with different results as to whether GR is preserved or discrepancies found. The goal of our future mission will be to use new, significantly deeper, spectroscopic and imaging survey datasets to extend these tests to higher source densities and more distant redshifts and to deliver the best dataset to constrain this ultimate test of gravity. \\

In addition to the cosmological probes, GAUSS will provide a legacy catalog to the community containing a significant fraction of the galaxies of our observable universe, about ten times more galaxies than the output of the imminent projects which are DESI, Vera Rubin observatory and Euclid, in spectrometry or in photometry. This catalog will be a unique database for multi-wavelength, multi-messenger studies. Different tracers of matter have different biases, which means different relation between the tracer and the underlying dark matter, and their combination can help lowering the cosmic variance, e.g. \cite{2013MNRAS.432..318A}. So the GAUSS catalog will have to be correlated with external probes, like weak lensing of CMB or neutral hydrogen intensity maps, among others, for consistency checks on one hand and to get still more robust and tighter constraints on cosmology on the other hand.\\

\section{Making progresses on matter distribution on large-scale in the Universe     }
\label{progresses}
From the observational point of view, although the origin of dark energy is unknown, its presence is effective in the recent universe (at redshift smaller than 5) and dominant in the very recent universe (at redshift smaller than typically 0.5). The presence of dark energy modifies the expansion rate, the growth rate of structure in the linear regime but also leads to subtle differences in the dynamics in the mildly nonlinear regime of structure formation. Given our lack of understanding on the origin of the cosmic acceleration, it is vital to obtain measurements of observational quantities that are sensitive to its presence and properties: the history of the expansion rate $H(z)$, the angular distance $D_A(z)$,  the growth rate of cosmic structure  $f_g(z)$, and diagnostics of gravity in the mildly nonlinear regime, like the $E_g$ quantity.\\
   
Several ground and space experiments are  already scheduled for these types of studies that will deliver their ultimate constraints at the 2030-2035  horizon. A metric of their performance for intercomparison has been proposed some years ago through the figure of merit (FoM), that is the inverse of the area of the constraint contours in the two parameters dark energy model Chevallier–Polarski–Linder (CPL). Although the full performance of a project is not  reduced to a single number, the FoM is still a very useful number for comparison. Here we used as much as possible the FoM computed in a flat cosmological model.  We have performed a forecast study  and establish a number of results that serve as a guideline for the present White Paper: individual probes like  WL (from shear measurement) and the 3D power spectrum (from spectroscopic samples) when combined between them provide a FoM much higher than that obtained on each probe individually, i.e. one plus one is (much) more than two. For instance, we have considered a fiducial survey with 30 gal/arcmin$^2$ for the photometric/WL part, and 1 gal/arcmin$^2$ for the spectroscopic part covering the redshift range $ 0.6 - 2$ over a sky area of 20,000 sq deg. With this setting one gets from the photometric clustering of galaxies (GCp) a FoM of 120, from the WL probe  a FoM of 60  and from  the spectroscopic probe a FoM of 300. A remarkable improvement is obtained by the addition of the cross-correlation between WL and GCp denoted by XC: the combination of WL, GCp, GCs  and XC boosts the FoM above  1000. This is already above foreseen surveys (like Euclid or Rubin observatory) for which FoM is anticipated to be around 500, but might correspond to what the combination of these various surveys can provide by 2035. This also illustrates that the critical ingredient for the ultimate constraint on dark energy properties resides in the cross-correlation  between the WL probe and the photometric sample GCp. The role of cross correlation between GCs and other probes has not yet been investigated in detail but preliminary investigation that we performed shows that an increase of 50\% on the FoM is realistic. This additional gain is however not taken into account in the following. \\

Let us now examine what improvement can be expected. In order to gain over existing or foreseen surveys, the main avenue is to increase the total number of objects that are observed. The gain anticipated compared to our fiducial survey is presented in Figure 1 in which the FoM is computed when the density of objects is increased by some factor compared to the fiducial one. The Rubin observatory  targets a density number of 40 galaxies/arcmin$^2$, targeting a FoM of 600 (in flat models) similar to the 30 galaxies/arcmin$^2$ Euclid photometric survey targeting a FoM of 400 (in non-flat models using the combination with Euclid spectroscopic data). For the spectroscopic sample, Euclid targets around 50 million galaxies, a number comparable to what is anticipated from the DESI sample (with noticeably different selection rules allowing DESI to achieve a higher FoM $\sim$ 150 for the spectroscopic probe alone). For our fiducial model, the spectroscopic sample will be around 700 milions spectra. Improving the density of galaxies in the spectroscopic sample allows us to improve the FoM by taking into account more small-scale data for the spectroscopic sample, a strategy of less interest at fainter density. \\

\begin{figure}
\centering
\vspace{-2ex}
\includegraphics[width=.99\columnwidth]{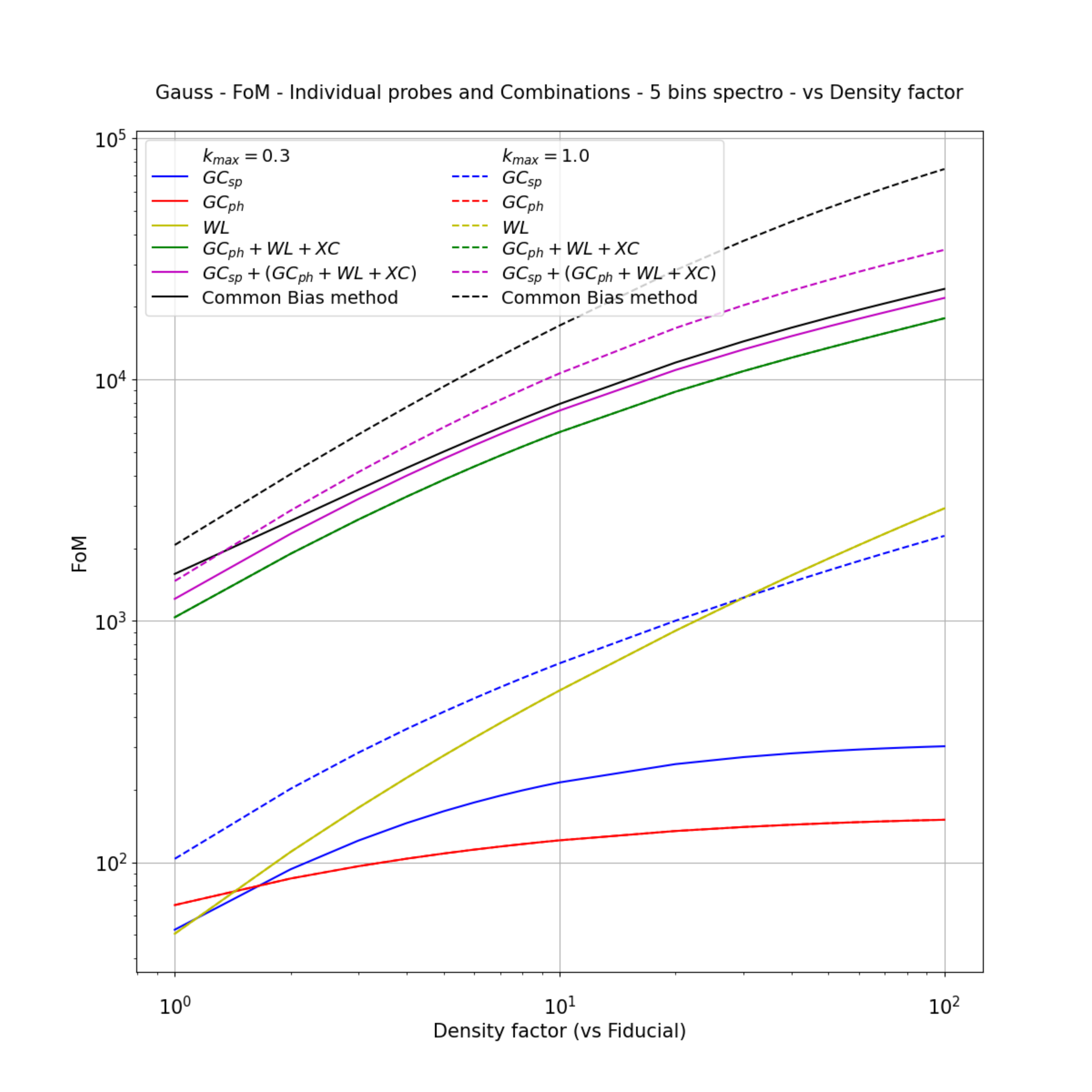}
\caption{An estimation of the gain compared to a fiducial survey (having 30 gal/arcmin$^2$ for the photometric part and 1 gal/arcmin$^2$ for the spectroscopic part, covering the redshift range $ 0.6 - 2$). Such a survey would achieve a FoM $\sim 2000$ higher than existing or foreseen surveys. This is also illustrative of what could be achieved by the combination of these various surveys (DESI, Euclid, Rubin Observatory, the Roman Space Telescope). With this strategy, as the density of objects being increased by some given factor (abscissa) one can monitor the gain in term of the FoM. Actually, GAUSS will have a larger redshift coverage, likely to lead to a higher FoM. Inclusion of more small-scale data for the spectroscopic sample ($k_{max} = 1$ compared to $k_{max} = 0.3 $), using the cross-correlation between photo metric samples and spectroscopic samples improves the FoM by a factor 2-3. As one can see spectroscopic galaxy clustering (GCs) and and photometric galaxy clustering (GCp) alone do not gain much from an increase of the density of  targets (with $k_{max} = 0.3 $). However, when combined with weak lensing and using XC the FoM increases by a factor nearly 10 reaching a value close to 10 000. Additional combination with the CMB might raise this number even more.}
\end{figure}

As shown in Figure 1, the final FoM, resulting from the combination of the main probes,  increases by a factor of more than ten relative to foreseen surveys if the number of targets in both the photometric and the spectroscopic samples are increased by a factor of 10. Increasing the number of targets by a further factor of ten would however not produce the same gain, instead the gain would be closer to a factor 3. Thus, this is a good indication that there is a significant gain to be achieved by increasing the number of targets by 10, but that the gain tends to weaken beyond that and would be extremely challenging as it would require deeper imaging than the Hubble ultra deep field. \\

The primary  gain anticipated from  a given telescope is proportional to the square of its diameter multiplied by the field of view. However, with a density of 300 objects per arcmin$^2$, for a ground-based telescope,  a significant fraction of galaxies in the field will overlap due to  the blurring produced by the atmosphere. This would make the weak lensing analysis extremely complex and challenging. Thus, only a space mission could provide accurate images of the quality necessary for weak lensing analyses if the number of objects has to be increased significantly. The limitations of ground-based image quality are compounded by the further limit on the availability of color information from the ground.  Broad optical as well infra-red photometry is necessary nowadays to estimate photometric redshifts and thereby optimally perform weak lensing analysis. Indeed, contamination by unresolved sources is a dominant limitation of weak lensing analysis that exquisite image quality has the potential to circumvent. In order to reach the foreseen density of objects, it is necessary to go 2.5 magnitudes deeper than the depth of typical working density of Euclid or the Rubin Observatory  for photometry and 4.5 magnitudes deeper for the spectroscopic survey. Compared to the Euclid space telescope this needs an increase in efficiency of around 60.\\

Spectroscopic surveys of galaxies from the ground are limited by telluric lines. Slitless spectroscopy is limited by the systematics intrinsic to this method. In order to achieve a  dense high-redshift  (in the range 1-4) spectroscopic galaxy sample the most efficient approach is infrared-space-based slit spectroscopy in the range 0.5-5 $\mu$m. \\


For dark energy studies, the final FoM is severely dependant on the combination of  WL and GC data. A survey focussed on the imaging part, with an exquisite quality will therefore achieve a remarkable progress on the FoM even without the combination of the spectroscopic part. This conclusion is however limited by the fact  that we do not have yet a full quantitative  investigation of the role of the cross-correlation between spectroscopic data and weak lensing data, and by the fact that on mildly non-linear scales the test of modified gravity theories needs  information coming from spectroscopic data. Identically, a large off-axis telescope is a technology that has not yet been implemented in space. A more classical solution could be adopted. \\

\section{The scientific landscape on the 2035 horizon}
\label{landscape}
The currently known landscape is summarized in Figure 2. It gathers the main space missions from Europe (mainly ESA but also DLR for eROSITA), NASA and Asia (which stands for JAXA and CSA projects). Most of these NASA, JAXA and CSA missions are also ESA Missions of Opportunity (MoOs). The black boxes outline missions that have cosmology among their main goals. The unique European mission, Euclid, should end around 2028. The current proposal concerns the next generation, after Euclid, Spherex and the Roman Space Telescope (previously known as WFIRST), eventually completed by the  spectroscopic ATLAS follow-up.  \\

\begin{figure}
\centering
\includegraphics[width=.99\columnwidth]{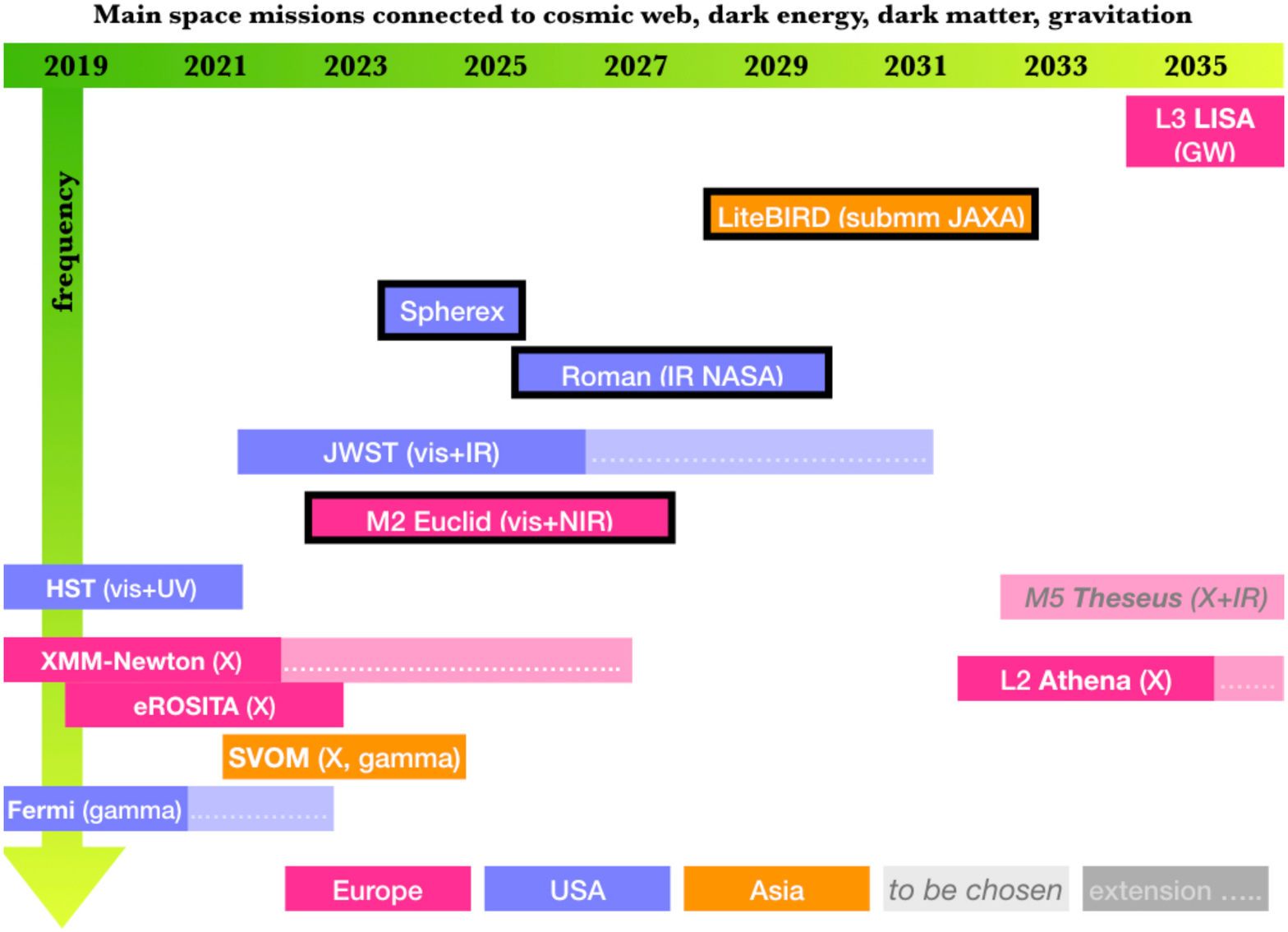}
\caption{ Main Space missions connected to cosmic web dark energy or gravitation}
\end{figure}

While Euclid should explore up to a few billion galaxies, with photometric redshifts, typically up to a redshift of 2 with spectroscopic measurements for a few percent of them; the Roman Space Telescope  should perform a fairly similar survey but deeper. In the meantime, Spherex will map the entire sky at low spatial resolution with a mid-spectral resolution in infrared. The goal of the GAUSS project is to obtain spectroscopic redshifts as well as the shapes of tens of billions of galaxies up to a redshift of ~5. \\

Observational cosmology takes place on the ground too. Here we summarize some of the main ground-based galaxy surveys. The Dark Energy Spectroscopic Instrument (DESI) is installed at the focal plane of a 4-meter telescope in Arizona to measure the spectra of more than 30 million galaxies and quasars covering 14,000 square degrees with the survey expected to start in 2020 \cite{Aghamousa:2016zmz}, now stopped because of the epidemic. The ESO cosmology redshift survey with the 4-meter Multi-Object Spectroscopic Telescope (4MOST) in Chile will start in 2022 \cite{2019Msngr.175...50R}, providing high resolution spectra of 8 million objects covering 15,000 square degrees up to $z=3.5$. The LSST (Large Synoptic Survey Telescope), recently renamed the Vera C. Rubin Observatory, will conduct a photometric survey, starting at the end of 2022, of $\sim 20,000$ square degrees with an 8-meter telescope in six bands based in Chile over ten years with a cadence suited to detect about 500 supernovae per night \cite{2019ApJ...873..111I}. While they have many advantages and will undoubtedly produce high quality science, none of these surveys is able to recover the shapes of galaxies with the same accuracy as space-based instruments, owing to the presence of the atmosphere. \\

The entire cosmic web, from the end of the epoch reionization until now, should be ultimately mapped in the visible/infrared domains. It would provide legacy data, thanks to the accuracy and the completeness of the survey, which can be correlated with CMB lensing, Sunyaev-Zeldovich or HI maps for instance. \\

Cosmological use of probes like WL and GC is more efficient when combined with cosmic microwave background (CMB) observations which directly map the first billion years. Some degeneracies are broken, for instance between the amplitude of primordial energy fluctuations and the optical depth linked to the reionization period, which currently prevent better constraints on the total mass of the neutrinos. The GAUSS project is perfectly timed to take advantage of the results of LiteBIRD and the S4 ground-based efforts. The cosmological parameters derived by these S4 experiments should be obtained with errors two times  lower than current experiments and a significantly more precise map of the CMB lensing, obtained by the large ground-based telescopes, should be available for correlation with the huge GAUSS galaxy catalogue. \\

The cosmological landscape will soon contain the Square Kilometre Array (SKA) which will conduct a huge spectroscopic galaxy survey, by detecting the 21 cm emission line of neutral hydrogen (HI) from around a billion galaxies over 3/4 of the sky, out to a redshift of $z \sim  2$ \cite{2015MNRAS.450.2251Y}. The survey should start by the end of the 2020s, ideally complementing the GAUSS galaxy survey.
Moreover, gravitational-wave astronomy, with the Einstein Telescope having succeeded  LIGO/Virgo on the ground and LISA in the sky, should be flourishing. The main synergies are the test of the law of gravitation at large scales or in the strong field regime, the possibility of testing the weak and strong equivalence principle, and also the measurement of the Hubble parameter today by using standard rulers or standard sirens, from telescopic and interferometric observations respectively.  Additionally, thanks to the Athena mission, the baryonic part of the cosmic web in the form of hot gas should be much better understood than now. Furthermore, the evolution of massive black holes and their role in galaxy evolution should be properly comprehended. So the knowledge of the baryonic ingredients of the cosmic web, as well as gas in galaxies, will be much more detailed than today, leading to a significant decrease of the systematic errors due to astrophysics. \\

\begin{figure}{h!}
\centering
\includegraphics[width=.99\columnwidth]{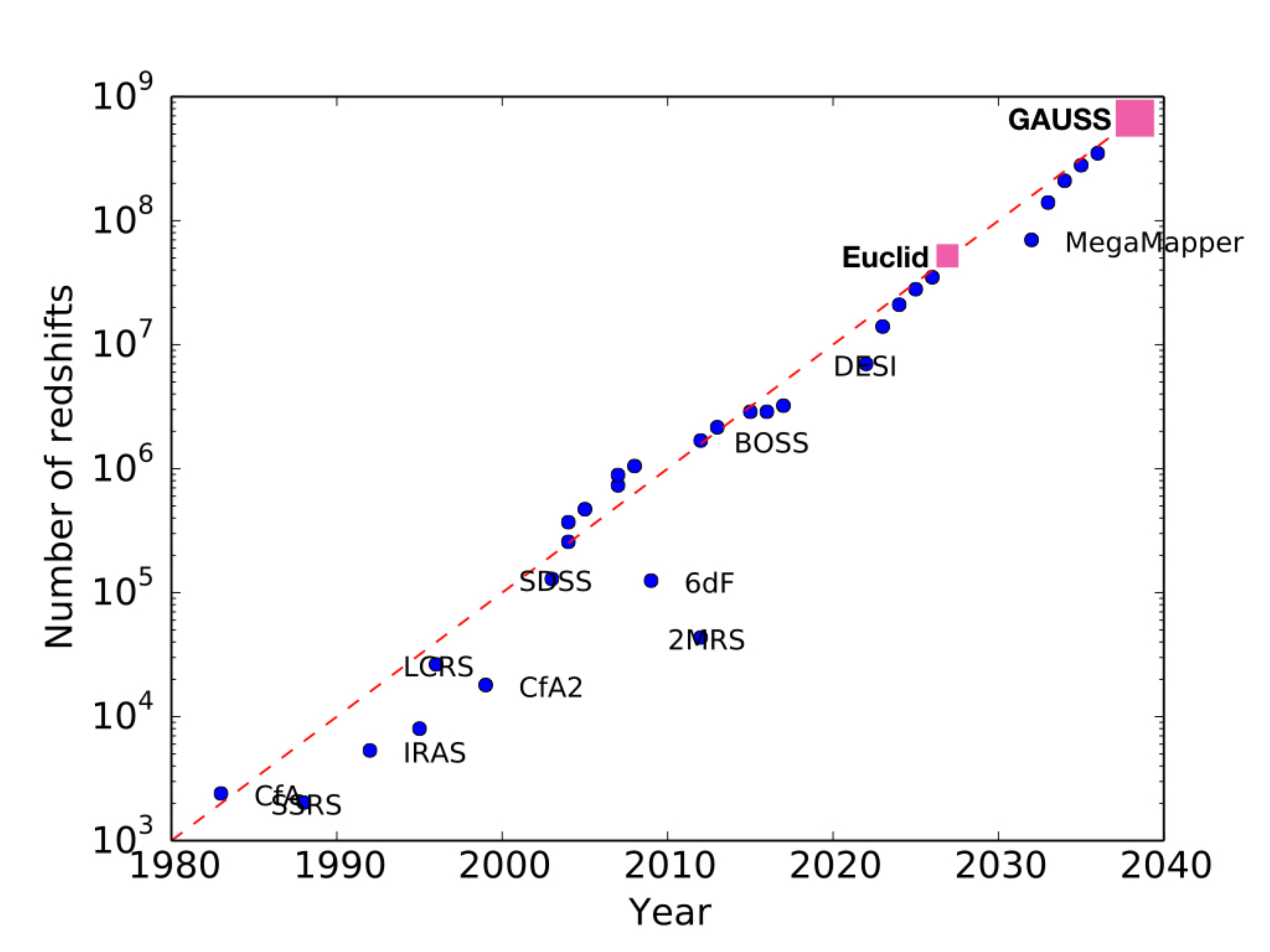}
\caption{Number of spectroscopic redshifts in surveys as a function of the year, extracted from the White Paper proposing MegaMapper, an answer to the Decadal Survey on Astronomy and Astrophysics Astro2020 call \cite{2019BAAS...51g.229S}. The Euclid space mission and the present GAUSS concept mission have been added for comparison.
}
\end{figure}

So, even if we cannot presume our knowledge of the dark energy after the Euclid era, we can be confident in the fact that our ability to measure the properties of dark energy and the neutrino mass will significantly increase in the era of Voyage 2050.

\section{Technological challenges}
\label{techno}
The proposed imaging survey should have color images in several bands. Although one could envision having single detectors allowing spectral information per pixel, here we consider a strategy where 8 bands per field are necessary, with a dichroic that would mean two focal planes with 4 bands each from 0.5 $\mu$m to 5 $\mu$m, those bands being achieved through filters on the detectors. This is aimed at avoiding the use of mechanical systems as much as possible.\\

The primary limitation will come from the total number of pixels of the camera: an appropriate sampling to fully benefit from the space image quality is to have a pixel size of 0.05 arcsec (although a pixel size of 0.1 arcsec would be acceptable). The field of view would be 2 degrees in size, to get an area of around 4 square degrees, 8 times wider than Euclid and slightly less than half the field of view of the Rubin Observatory. Assuming a pixel size of 5  $\mu$m and a ten Giga-pixels camera, the physical size will be around 70 cm. Assuming half of the field is used by the imaging system with 8k$\times$8k detectors, this needs 160 detectors in the focal plane. An extra factor of efficiency of ~8 is needed which implies the need to have a 3-4 meter class telescope. In order to get the highest image quality for the images, an off-axis mirror will be the preferred solution. Such solutions have been investigated and their advantages have been quantified \cite{2010SPIE.7731E..1GL, 2014PASP..126..386S}. \\

In order to achieve a massive spectroscopic sample we propose to use Digital Micro-­mirror Devices as slit selectors \cite{2009ExA....23...39C}. Although this technology has not yet been demonstrated in space, it has been used in astronomy \cite{2006SPIE.6269E..15M} and we anticipate that this or some equivalent technology  will allow massive slit spectroscopy on selected targets. A significant fraction (50\% as a guideline) of the focal plane should be used for this. Sharing the focal plane in this way will avoid having moving mechanical systems for dealing with both spectroscopic and photometric channels. Spectroscopic devices will be located in the outer part of the field as the spatial resolution is less critical for spectroscopy. 16 4k$\times$4k detectors with 10$\mu$m pixels can achieve a mutiplex factor greater than 20 000. Using slit spectroscopy will suppress the sky background inherent in the slitless-grism spectroscopy usually used in space missions (HST and Euclid for example) thus enabling a significant gain in sensitivity.\\

\section{Conclusion}
\label{conclusions}
The statistical distribution of matter over a very large volume of the universe will remain the primary tool to investigate the source of the accelerated expansion of the universe as well as the physics of the very early universe. Large photometric and spectroscopic surveys of galaxies over the same sky area are particularly efficient for these objectives. A space mission like GAUSS will surpass by more than an order of magnitude all currently foreseen projects (Euclid, Rubin Observatory, DESI, Roamn Space Telescope, …)  thanks to a very deep flux limit (including in the infrared domain up to $\sim 5\mu$m), a very high multiplexing capability allowing to map the distribution of a unique tracer from redshift 0.5 to 5, with limited systematics, and the power of probe combinations and their cross-correlation within a single experiment, a unique advantage of GAUSS over most existing projects (with the exception of Euclid). This would allow, for instance, a definitive measurement of the total mass of neutrinos from a single experiment, and provide major progress in our understanding of Dark Energy and Inflation, two major problems of both cosmology and fundamental physics. Thus, a mission like GAUSS linking the early and late phases of cosmic evolution, with their hugely different energy scales, provides unique clues for cosmology, gravitation, and inflation physics, without any equivalent tool for investigating these topics.


%
%


\bibliographystyle{plain}
\bibliography{Bibliography}

\end{document}